\documentclass[10pt]{article}

\usepackage{%color,
graphicx,amsmath,amssymb,bm}

\newtheorem{prop}{Prop}[section]
\newtheorem{lemma}[prop]{Lemma}

\newtheorem{theorem}[prop]{Theorem}

\newtheorem{corollary}[prop]{Corollary}

\begin{document}

\title{
Zero-variance of perturbation Hamiltonian density
in perturbed spin systems%\subtitle{Do you have a subtitle?\\ If so, write it here
}
\author{C. Itoi
\\
Department of Physics,   
GS $\&$ CST, 
Nihon University
}

\maketitle
\begin{abstract}  We study effects of perturbation Hamiltonian to quantum spin systems which can include quenched disorder.   
 Model-independent inequalities are derived, using an additional artificial disordered perturbation. These inequalities enable us to
 prove  that  the variance of the perturbation Hamiltonian density
vanishes in the infinite volume limit even if the artificial perturbation is switched off.  
This  theorem  is applied to spontaneous symmetry breaking phenomena in a disordered classical spin model, 
a quantum spin model without disorder and a disordered quantum spin  model. 
\end{abstract}

\section{Introduction}
We study quantum spin systems on a lattice
$V_N$ which consists of $N$ sites.  
%One of the most important example is given by nearest neighbor bonds
%$B_L=\{\{ x,y\}|x,y \in V_N, |x-y|=1 \}$.
A spin operator  $S^{p}_j$ $(p=x,y,z)$ at  a site $j \in V_N$
on a Hilbert space ${\cal H} :=\bigotimes_{j \in V_N} {\cal H}_j$ is
defined by a tensor product of the spin matrix acting on ${\cal H}_j \simeq {\mathbb C}^{2S+1}$ and unities, where $S$ is an arbitrary fixed half integer.
These operators are self-adjoint and satisfy the commutation relations
 \begin{equation}
[S_j^x,S_k^y]=i \delta_{j,k} S_j^z ,  \ \  \ \ \  [S_j^y,S_k^z]=i \delta_{j,k}  S_j^x, \ \ \  \ \ 
[S_j^z,S_k^x]=i \delta_{j,k} S_j^y,
 \end{equation}
and the spin at each site $i \in V_N$ has a fixed magnitude 
 \begin{equation}
 \sum_{a=x,y,z}(S_j^a)^2 =S(S+1) {\bf 1}.
 \end{equation}
We define an  unperturbed Hamiltonian $H_N({\bf S})$ first.
 ${\cal P} (V_N)$ denotes the collection of all subsets of $V_N$.
 Let $C_N \subset {\cal P}(V_N)$ be a collection of interaction ranges and 
 let ${\bf J} = (J_X)_{X \in C_N}$ be a sequence of  real valued random variables with a finite  expectation ${\mathbb E} |J_X |$ for each $X \in C_N$,
where ${\mathbb E}$ denotes the expectation over ${\bf J}$ for its functions.  ${\bf S}_X$ denotes a sequence of spin operators  $(S_j^p)_{j \in X, p=x,y,z}$  on a subset $X$ and $\varphi$ is a self-adjoint valued 
 function of ${\bf S}_X$.
 We consider a model defined by the following Hamiltonian with $C_N$, $\varphi$ and ${\bf J}$
\begin{equation}
H_N({\bf S},{\bf J}) := \sum_{X\in C_N} J_X \varphi( {\bf S}_X). \label{unpert}
\end{equation}
%and there exist a positive number $C$ and $C'$ independent of $N$, such that
%\begin{equation}
%{\marthbb E} \| H_N({\bf S}) \| \leq C N \ \ \  and 
%\end{equation}

One can assume a symmetry of the Hamiltonian $H_N({\bf S},{\bf J})$, if one is interested in symmetry breaking phenomena.
To detect a spontaneous symmetry breaking, long-range order of 
order operator  $ h_N({\bf S})$ is utilized  in the symmetric Gibbs state.  Although the symmetric Gibbs state with long-range order is mathematically well defined, 
such state is unstable due to strong fluctuation and it cannot be realized.  
On the other hand,  it is believed  that a perturbed Gibbs state with infinitesimal
symmetry breaking Hamiltonian %$H_I({\bf S})$ 
is stable and  realistic.
Consider a perturbed Hamiltonian as a function of spin operators ${\bf S} = (S_j^p)_{j \in V_N, p=x,y,z}$
\begin{equation}
H:= H_N({\bf S}, {\bf J})-N\lambda h_N({\bf S}),
\label{hamil}
\end{equation}
%Hereafter, we use lighter notation $H$ instead of $H({\bf S}, {\bf J}, g)$.
%Consider the following operator 
%\begin{equation}
%H_I({\bf S},\lambda)=- N  \lambda   h_N({\bf S}),
%\label{per}
%\end{equation}
where $h_N({\bf S})$ is a bounded operator and  $\lambda \in {\mathbb R}$.
Assume upper bounds on $h_N({\bf S})$
\begin{equation}
\| h_N({\bf S}) \| \leq C_h,   
\end{equation} 
where $C_h$ is a positive constant independent of the 
system size $N$. The operator norm is defined by 
$\| O \|^2 :=\sup_{(\phi,\phi)=1 } (O \phi, O \phi)$ for an arbitrary linear operator $O$ on ${\cal H}$.
To study spontaneous symmetry breaking, one can regard $h_N({\bf S})$ as an order operator which breaks the
 symmetry. 
 For instance, $h_N$ is a
spin density
 \begin{equation}
h_N({\bf S}) = \frac{1}{N} \sum_{j \in V_N} S_j^z.
 \end{equation}

Define  Gibbs state with the Hamiltonian (\ref{hamil}).
For  $\beta>0$,  the  partition function is defined by
\begin{equation}
Z_N(\beta,\lambda, {\bf J}) := {\rm Tr} e^{ - \beta  H_N({\bf S}, {\bf J})+\beta N\lambda h_N({\bf S})}
\end{equation}
where the trace is taken over the Hilbert space ${\cal H}$.
Let  $f$ be an arbitrary function 
of spin operators.  
The  expectation of $f$ in the Gibbs state is given by
\begin{equation}
\langle f({\bf S}) \rangle_{\lambda}=\frac{1}{Z_N(\beta, \lambda, {\bf J})}{\rm Tr} f({\bf S})  e^{ - \beta  H_N({\bf S}, {\bf J})+\beta N\lambda h_N({\bf S})}.
\end{equation}
Define the following function from the partition function
 \begin{equation}
\psi_N(\beta, \lambda,{\bf J}):=\frac{1}{N} \log Z_N(\beta,\lambda, {\bf J}),
 \end{equation}
and its expectation 
 \begin{equation}
p_N(\beta, \lambda):={\mathbb E} \psi_N(\beta, \lambda, {\bf J}).
 \end{equation}
where ${\mathbb E}$ denotes the expectation over ${\bf J}$.
The function $-\frac{N}{\beta} \psi_N$ is called free energy of the sample in statistical physics.\\

In the present paper, we require  the following three assumptions on the Gibbs state defined by the perturbed  Hamiltonian (\ref{hamil}). \\
 
\noindent
{{\bf Assumption 1}
  \it
The  infinite volume limit  of the function $p_N$ 
 \begin{equation}
 p(\beta, \lambda) = \lim_{N\nearrow\infty}   p_N(\beta, \lambda), 
 \end{equation}
exists for each $(\beta, \lambda) \in (0,\infty) \times {\mathbb R}$.}\\
%\noindent
%{{\bf Assumption 3}\itThe infinite volume limit  of the expectation of an arbitrary bounded function $f$ of spin operators 
% \begin{equation} \lim_{N\nearrow\infty}   {\mathbb E} \langle f({\bf S}) \rangle _{\lambda,\mu, u},  \begin{equation}
%exists for each $(\beta, \lambda,\mu, u) \in (0,\infty) \times {\mathbb R}^{2} \times [0,1]$.}\\

\noindent
{\it {\bf Assumption 2} The  variance of 
$ \psi_N$ vanishes in the infinite volume limit 
$$
\lim_{N\nearrow\infty}{\mathbb  E} [\psi_N(\beta, {\bf J},  \lambda)- p_N(\beta, \lambda) ]^2=0,% \leq \frac{C}{N},
$$
for each $(\beta, \lambda) \in (0,\infty) \times {\mathbb R}$
 }\\

\noindent
{{\bf Assumption 3} \it  The following commutator of
the perturbation operator $h_N$ and the Hamiltonian vanishes in the infinite volume limit %satisfies the following  bound 
 \begin{equation}
\lim_{N\nearrow\infty} \|[h_N({\bf S}), [H{(\bf S}, {\bf J}), h_N({\bf S})]]\| =0,
 \end{equation}
   for an arbitrarily fixed sequence $\bf J$.
}\\

In the present paper, we prove the following main theorem for an arbitrary spin model with 
the Hamiltonian (\ref{hamil})  satisfying Assumptions 1, 2 and 3. \\

{\theorem \label{MT} 
Consider  a quantum spin model defined by the Hamiltonian
(\ref{hamil}) satisfying Assumptions 1, 2 and 3.
 The expectation of the perturbation operator
 \begin{equation}\displaystyle{\lim_{N\nearrow\infty }{\mathbb E}  \langle h_N({\bf S}) \rangle_{\lambda}}, 
 \end{equation}
 exists  in the infinite volume limit  for almost all $\lambda$ and 
 its variance in the Gibbs state  and the distribution of disorder vanishes 
 
\begin{equation}
\lim_{N \nearrow \infty }{\mathbb E}\langle ( h_N({\bf S})-{\mathbb E} \langle h_N({\bf S}) \rangle_{\lambda} )^2 \rangle_{\lambda}=0,
\end{equation} 
in the infinite volume limit  for almost all $\lambda \in {\mathbb R}$.}\\

Theorem \ref{MT} implies  also the existence of the following infinite volume  limit for almost all $\lambda \in {\mathbb R}$
 \begin{equation}
\lim_{N\nearrow\infty}
  {\mathbb E} \langle {h_N({\bf S})}^2 \rangle_{\lambda}=(\lim_{N\nearrow\infty}{\mathbb E} \langle {h_N({\bf S})}\rangle_{\lambda})^2.
 \end{equation}
The perturbation operator $h_N$ is self-averaging  in the perturbed model.  
Although the claim of Theorem \ref{MT} is physically quite natural and physicists believe it by
their experiences supported by lots of examples,  it has never been proved rigorously in general setting.
In section 2, we prove Theorem \ref{MT}.  In  section 3, 
we apply Theorem \ref{MT} to  spontaneous symmetry breaking phenomena in several examples.

\section{Proof}
In this section, we introduce an extra perturbation Hamiltonian with 
a  quenched disorder to  prove Theorem \ref{MT}.  
Consider the following perturbed Hamiltonian
\begin{equation}
H=H_N({\bf S}, {\bf J})- (N  \lambda +N^\alpha \mu g) h_N({\bf S}).
\label{per}
\end{equation}
where  $g$ is a standard Gaussian random variable with 
$\mu \in {\mathbb R} $, and $\alpha \in (0,1)$. 
%We choose an exponent $\alpha > 0$ to take the infinite volume limit
%after evaluations of physical quantities depending on the unperturbed Hamiltonian $H_N$.
The introduced random variable $g$ is artificial and our final goal %purpose 
is to study  the model  at
$\mu=0$.
In this section, the symbol ${\mathbb E}$ denotes the expectation over all random variables ${\bf J}, g$.
and ${\mathbb E}_g $ denotes that over only $g$.
In this section,  we regard
the following  functions for  the Hamiltonian (\ref{per}) 
\begin{equation}
Z_N(\beta,\lambda, {\bf J}, \mu g) := {\rm Tr} e^{ - \beta H_N({\bf S}, {\bf J})+\beta (N  \lambda +N^\alpha \mu g) h_N({\bf S})},
\end{equation}
and 
 \begin{equation}
\psi_N(\beta, \lambda,{\bf J}, \mu g):=\frac{1}{N} \log Z_N(\beta,\lambda, {\bf J}, \mu g),
 \end{equation}
as  functions of $(\beta,\lambda, {\bf J}, \mu g)$
 The expectation of $\psi_N$ is
 \begin{equation}
p_N(\beta, \lambda, \mu):={\mathbb E} \psi_N(\beta, \lambda, {\bf J}, \mu g).
 \end{equation}
%where ${\mathbb E}$ denotes the expectation over ${\bf J}$ and $g$.
%where the trace is taken over the Hilbert space ${\cal H}$.
For  an arbitrary function  $f$, 
of spin operators.  
the Gibbs  expectation of $f$ is 
\begin{equation}
\langle f({\bf S}) \rangle_{\lambda,\mu g}=\frac{1}{Z_N(\beta, \lambda, {\bf J},  \mu g)}{\rm Tr} f({\bf S})  e^{ - \beta H_N({\bf S}, {\bf J})+\beta (N  \lambda +N^\alpha \mu g) h_N({\bf S})}.
\end{equation}
\\

Here, we introduce a fictitious time  $t \in [0,1]$ and define a time evolution of operators with the Hamiltonian.
Let $O$ be an arbitrary self-adjoint operator, and we define an operator valued function  $O(t)$ of $t\in[0,1]$  by
\begin{equation}
 O(t):= e^{-tH} O  e^{tH}.
\end{equation}
Furthermore, we define the  Duhamel product of time independent operators $O_1,O_2, \cdots, O_k$
%$ O_1(t_1),  \cdots,  O_k(t_k)$  
by
 \begin{equation}
( O_1,O_2, \cdots, O_k)_{\lambda,\mu g} :=\int_{[0, 1]^k} dt_1\cdots dt_k \langle {\rm T}[ O_1(t_1)  O_2(t_2) \cdots  O_k(t_k) ]\rangle_{\lambda,\mu g},
 \end{equation}
where the symbol ${\rm T}$ is a multilinear mapping of the chronological ordering.
If we define a partition function with arbitrary self adjoint operators  $O_1, \cdots, O_k$ and real
numbers $x_1, \cdots, x_k$
 \begin{equation}
Z(x_1,\cdots, x_k) := {\rm Tr} \exp \beta \left[-H+\sum_{i=1} ^k x_i O_i \right],
 \end{equation}
the Duhamel product of $k$ operators represents
 the $k$-th order derivative of the partition function 
 \cite{Cr,GUW,S}
 \begin{equation}\beta^k( O_1,\cdots, O_k)_{\lambda, \mu g}=\frac{1}{Z}
\frac{\partial ^k Z}{\partial x_1 \cdots \partial  x_k}.
 \end{equation}
Furthermore,  a truncated Duhamel product is defined by 
 \begin{equation}\beta^k( O_1 ; \cdots ;   O_k)_{\lambda, \mu g}=
\frac{\partial ^k }{\partial x_1 \cdots \partial  x_k} \log Z.
 \end{equation}

The following lemma can be shown in the standard convexity argument to obtain the Ghirlanda-Guerra identities  \cite{AC,CG2,GG,I,I2,Pn,T}
in classical and quantum disordered systems. 
The proof can be done on the basis of of convexity  of functions $\psi_N$, $p_N$, $p$ and their almost everywhere differentiability and 
Assumptions 1 and 2.

{\lemma \label{Delta} For almost all $\lambda \in {\mathbb R}$, the infinite volume limit
\begin{equation}
\frac{\partial p}{\partial \lambda} (\beta, \lambda, 0) = \lim_{N\nearrow\infty} \beta {\mathbb E} \langle h_N({\bf S}) \rangle_{\lambda,0} 
\label{lim1}
\end{equation}
exists  and the following variance vanishes
\begin{equation}
\lim_{N\nearrow\infty} [{\mathbb E} \langle h_N({\bf S}) \rangle_{\lambda,0}^2 -({\mathbb E} \langle h_N({\bf S})\rangle_{\lambda,0}) ^2]=0,
\end{equation}
for each $\beta \in (0,\infty)$.\\

\noindent
Proof.}
In this proof, regard $p_N(\lambda)$ $p(\lambda)$ and $\psi_N(\lambda)$ as functions of $\lambda$ for lighter notation. 
Define the following functions 
\begin{eqnarray}
&&w_N(\epsilon) := \frac{1}{\epsilon}[|\psi_N(\lambda+\epsilon )-p_N(\lambda+\epsilon)|+|\psi_N(\lambda- \epsilon)-p_N(\lambda-\epsilon)|
+|\psi_N(\lambda )-p_N(\lambda)| ]\nonumber 
\\
&&e_N(\epsilon ):=\frac{1}{\epsilon}[|p_N(\lambda+\epsilon )-p(\lambda+\epsilon)|+|p_N(\lambda- \epsilon)-p(\lambda-\epsilon)|
%\nonumber \\&&
+|p_N(\lambda )-p(\lambda)|],%\nonumber
\end{eqnarray}
for $\epsilon > 0$.
Assumption 1 and Assumption 2 on $\psi_N$  give
\begin{equation}
\lim_{N\nearrow\infty} {\mathbb E}w_N(\epsilon)=0, \ \ \  \lim_{N\nearrow\infty} e_N(\epsilon)=0,% \leq \frac{3}{\epsilon} \sqrt{\frac{C}{|V_N|}},
\end{equation}
for any $\epsilon > 0$. 
Since $\psi_N$, $p_N$ and $p$ are convex functions of $\lambda$, we have
\begin{eqnarray} 
&&\frac{\partial \psi_N}{\partial \lambda}(\lambda) - \frac{\partial  p}{\partial \lambda}(\lambda)% \nonumber \\&&
\leq \frac{1}{\epsilon} [\psi_N(\lambda+\epsilon)-\psi_N(\lambda)]- \frac{\partial  p}{\partial \lambda}%\nonumber 
\\
&&\leq \frac{1}{\epsilon} [\psi_N(\lambda+\epsilon)-p_N(\lambda+\epsilon)+p_N(\lambda+\epsilon)-p_N(\lambda)
+p_N(\lambda)-\psi_N(\lambda) \nonumber \\
&& -p(\lambda+\epsilon) +p(\lambda+\epsilon)+p(\lambda)-p(\lambda) ]- \frac{\partial  p}{\partial \lambda}(\lambda) %\nonumber 
\\
&&\leq \frac{1}{\epsilon} [ |\psi_N(\lambda+\epsilon)-p_N(\lambda+\epsilon)|
+|p_N(\lambda)-\psi_N(\lambda)| +|p_N(\lambda+\epsilon)-p(\lambda+\epsilon)|\nonumber 
\\ 
&&+|p_N(\lambda)-p(\lambda)| ]+\frac{1}{\epsilon}[ p(\lambda+\epsilon)-p(\lambda)] - \frac{\partial  p}{\partial \lambda}(\lambda) %\nonumber
 \\
&&\leq w_N(\epsilon) +e_N(\epsilon)
+  \frac{\partial p}{\partial \lambda}(\lambda+\epsilon) - \frac{\partial  p}{\partial \lambda}(\lambda). % \nonumber
\end{eqnarray}  
As in the same calculation, we have
\begin{eqnarray} 
&&\frac{\partial \psi_N}{\partial b\lambda}(\lambda) - \frac{\partial  p}{\partial \lambda}(\lambda) 
\geq \frac{1}{\epsilon}[\psi_N(\lambda)-\psi_N(\lambda-\epsilon)] - \frac{\partial  p}{\partial \lambda}(\lambda) %\nonumber 
\\&&
\geq -w_N(\epsilon) -e_N(\epsilon)+ \frac{\partial p}{\partial \lambda}(\lambda-\epsilon)- \frac{\partial  p}{\partial \lambda}(\lambda) . %\nonumber 
\end{eqnarray}  
Then, 
\begin{eqnarray} 
{\mathbb E}\Big|\frac{\partial \psi_N}{\partial \lambda}(\lambda) - \frac{\partial p}{\partial \lambda}(\lambda)\Big| \leq {\mathbb E}w_N(\epsilon)
+e_N(\epsilon)+  \frac{\partial p}{\partial \lambda}(\lambda+\epsilon) -  \frac{\partial p}{\partial \lambda}(\lambda-\epsilon).%\nonumber
\end{eqnarray}  
Convergence of $p_N$ in  the infinite volume limit implies 
\begin{eqnarray}
&&\lim_{N\nearrow\infty } {\mathbb E}\Big| \beta \langle h_N ({\bf S}) \rangle_{\lambda,0} - \frac{\partial p}{\partial \lambda}(\lambda)\Big|%\nonumber \\&&
%=\beta \sqrt{u }\lim_{N\nearrow\infty }\lim_{M \rightarrow \infty } {\mathbb E}| \langle h_N  \rangle_{b,i} -{\mathbb E} \langle h_N\rangle_{b,i} |\nonumber \\&&
\leq  \frac{\partial p}{\partial \lambda}(\lambda+\epsilon) -  \frac{\partial p}{\partial \lambda}(\lambda-\epsilon),% \nonumber
\end{eqnarray}
The right hand side vanishes, since the convex function$p(\lambda)$ is continuously 
differentiable almost everywhere and $\epsilon >0$ is arbitrary. Therefore
\begin{equation}
\lim_{N\nearrow\infty}{\mathbb E} \Big| \beta  \langle h_N ({\bf S}) \rangle_{\lambda,0} - \frac{\partial p}{\partial \lambda}(\lambda)\Big|=0.
\label{limit3}
\end{equation}
for almost all $\lambda$.   Jensen's inequality gives 
\begin{equation}
\lim_{N\nearrow\infty}\Big| {\mathbb E}\beta  \langle h_N ({\bf S}) \rangle_{\lambda,0} - \frac{\partial p}{\partial \lambda}(\lambda)\Big|=0.
\end{equation}
 This implies the first equality (\ref{lim1}).  
%Regarding $p,p_N$ and $\psi_N$ as convex functions of $c$,the second equality (\ref{limit2}) is also obtained. 
Since the  $p(\lambda)$ is continuously differentiable almost everywhere in ${\mathbb R}$, 
these equalities imply also
 \begin{equation}
\lim_{N\nearrow\infty }  {\mathbb E} |\langle h_N ({\bf S}) \rangle_{\lambda,0} - 
 {\mathbb E} \langle h_N({\bf S})  \rangle_{\lambda,0}| =0.
 \end{equation}
The bound on $h_N({\bf S})$  concludes the following limit
 \begin{equation}
\lim_{N\nearrow\infty }  {\mathbb E} (\langle h_N ({\bf S}) \rangle_{\lambda,0} - 
 {\mathbb E} \langle h_N({\bf S})  \rangle_{\lambda,0})^2 \leq 
 2C_h \lim_{N\nearrow\infty }  {\mathbb E} |\langle h_N ({\bf S}) \rangle_{\lambda,0} - 
 {\mathbb E} \langle h_N({\bf S})  \rangle_{\lambda,0}|=0.
 \end{equation}
This  completes the proof.
$\Box$\\

Note that Lemma \ref{Delta} guarantees the existence of the following 
infinite volume limit for almost all $\lambda \in {\mathbb R}$
 \begin{equation}
\lim_{N\nearrow\infty}{ \mathbb E}\langle h_N({\bf S}) \rangle_{\lambda, 0}^2 
=(\lim_{N\nearrow\infty}{ \mathbb E}\langle h_N({\bf S}) \rangle_{\lambda, 0})^2
 \end{equation}

{\lemma \label{1}  
Let $f({\bf S})$ be a function 
of spin operators which is self-adjoint and  bounded by a constant $C_f$ independent of $N$
 \begin{equation}
\| f({\bf S})\| \leq C_f.
 \end{equation}
For any $(\beta, \lambda,\mu) \in [0,\infty) \times {\mathbb R}^{2}$, any positive integer
$N$ and $k$, arbitrarily fixed ${\bf J} $, 
an upper bound on the  following $k$-th order derivative 
 is given by 
\begin {equation}
\Big|{\mathbb E}_g \frac{\partial^k}{\partial \lambda^k} \langle f({\bf S}) \rangle_{\lambda,\mu g} \Big| \leq \sqrt{k! }C_f |\mu|^{-k} N^{k(1-\alpha)},
\end{equation}
where ${\mathbb E}_g$ denotes the expectation only over the standard Gaussian random variable $g$. \\

\noindent
Proof.}  
  Let $g,g'$ be i.i.d. standard Gaussian random variables and 
define a function with a parameter $u \in [0,1]$
 \begin{equation}
G(u) :=\sqrt{u} g+\sqrt{1-u} g'.
 \end{equation}
  Define a generating  function $\chi_f$ of the parameter $u \in [0,1]$ for $f$ by
\begin{equation}
\chi_f(u) :=
 {\mathbb E}_g  [{\mathbb E}_g' \langle f({\bf S}) \rangle_{\lambda,\mu G(u)}]^2,
\end{equation}
where ${\mathbb E}'_g$ is expectation over only $g'$ and ${\mathbb E}_g$ is expectation over only random variables $g$ and $g'$. 
 This generating function $\chi_f$ is  a generalization of a function introduced by Chatterjee \cite{C}.
First we prove the following  formula 
\begin{equation}
\frac{d^k \chi_f}{{d u}^k} (u) = N^{2(\alpha -1)k} \mu^{2k} 
{\mathbb E}_g \Big[{\mathbb E}_g'  \frac{\partial^k}{\partial \lambda^k}\langle f({\bf S}) \rangle_{\lambda,\mu G(u)} \Big]^2.
\label{gamma}
\end{equation}
The following inductivity for a positive integer $k$ proves this formula. \\
For $k=1$, the first derivative  of $\chi_f$ is 
\begin{eqnarray}
\chi_f'(u) &=& N^\alpha \beta \mu {\mathbb E}_g {\mathbb E}' _g\langle f({\bf S}) \rangle_{\lambda,\mu G(u)}
{\mathbb E}' _g\Big(\frac{g}{\sqrt{u}}-\frac{g'}{\sqrt{1-u}}\Big)( f({\bf S}) ; h_N({\bf S}))_{\lambda,\mu G(u)} \nonumber \\ 
&=& N^\alpha \beta \mu {\mathbb E}_g \Big[ \frac{1}{\sqrt{u}}\frac{\partial }{\partial g}
{\mathbb E}_g' \langle f({\bf S}) \rangle_{\lambda,\mu G(u)}{\mathbb E}_g'( f({\bf S}) ; h_N({\bf S}))_{\lambda,\mu G(u)} \nonumber \\
& &-{\mathbb E}_g' \langle f({\bf S}) \rangle_{\lambda,\mu G(u)}{\mathbb E}_g' \frac{1}{\sqrt{1-u}}\frac{\partial }{\partial g'}( f({\bf S}) ; h_N({\bf S}))_{\lambda,\mu G(u)} \Big]\\ 
&=&  N^{2\alpha} \beta^2 \mu^2 {\mathbb E}_g[
{\mathbb E}_g' ( f({\bf S}) ; h_N({\bf S}))_{\lambda,\mu G(u)}]^2 \\
&=& N^{2(\alpha -1)} \mu^{2} 
{\mathbb E}_g \Big[{\mathbb E}_g'  \frac{\partial}{\partial \lambda}\langle f({\bf S}) \rangle_{\lambda,\mu G(u)} \Big]^2,
\end{eqnarray}
where  integration by parts over $g$ and $g'$ has been used.  
If the validity of  the formula (\ref{gamma})  is assumed 
for an arbitrary positive integer $k$,  then (\ref{gamma}) for $k+1$ can be proved using integration by parts. 
The formula (\ref{gamma})  shows that $k$-th derivative of $\chi_f(u)$ is positive semi-definite for any $k$, 
%The Taylor series of the function $\chi_f(u)$ around  $u=0 $.
then it is a monotonically increasing function of $u$.
From Taylor's theorem, there exists $v\in (0,u)$ such that 
 \begin{equation}
\chi_f(u) =  \sum_{n=0}^{k-1}\frac{u^n}{n!} \chi_f^{(n)} (0) + 
 \frac{u^k}{k!} \chi_f^{(k)} (v).
 \end{equation}   
Each term in this series  is bounded from the above by 
 \begin{equation}
\chi_f(1) ={\mathbb E}_g \langle f({\bf S}) \rangle_{\lambda,\mu g}^2
\leq \| f \|^2 \leq C_f^2 .
 \end{equation}
The definition of $G(u)$ and the formula (\ref{gamma}) give
\begin{eqnarray}
N^{2(\alpha -1)k} \mu^{2k} \Big[
{\mathbb E}_g %{\mathbb E}'  
\frac{\partial^k}{\partial \lambda^k}\langle f({\bf S}) \rangle_{\lambda,\mu g} \Big]^2&=&
N^{2(\alpha -1)k} \mu^{2k}
{\mathbb E}_g \Big[{\mathbb E}_g'  \frac{\partial^k}{\partial \lambda^k}\langle f({\bf S}) \rangle_{\lambda,\mu G(0)} \Big]^2 \nonumber \\
&=& \frac{d^k \chi_f}{{d u}^k} (0) \leq k! \chi_f(1) \leq   k! C_f^2.
\end{eqnarray}
This completes the proof. $\Box$\\

{\lemma \label{cont0}
The infinite volume limit 
 \begin{equation}
p(\beta, \lambda, \mu):=\lim_{N \nearrow\infty} p_N(\beta, \lambda, \mu) 
 \end{equation} 
exists for  each $(\alpha, \beta, \lambda, \mu) \in (0,1) \times  (0,\infty) \times {\mathbb  R}^2$, 
and
$
p(\beta, \lambda, \mu)= p(\beta, \lambda,0).
$
\\

\noindent
Proof.}  Integration of the derivative function of $p_N$ 
over the interval $(0,\mu)$  and integration by parts with respect to $g$ give
\begin{eqnarray}
&&|p_N(\beta,\lambda,\mu) -p_N(\beta,\lambda,0) |
=\Big| \int_0^\mu dx \frac{\partial }{\partial x} p_N (\beta, \lambda,x)\Big| \\
&&=\Big| \int_0^\mu dx {\mathbb E } N^{\alpha-1} \beta g \langle h_N({\bf S})\rangle_{ \lambda, x g} \Big|\\
&&=\Big|\int_0^\mu dx {\mathbb E }N^{2\alpha-1}\beta^2 x (h_N({\bf S}) ; h_N({\bf S}))_{\lambda,x g} \Big| \\
&&= \Big| \int_0^\mu dx x N^{2\alpha-2} \beta \frac{\partial}{\partial \lambda}{\mathbb E }  \langle h_N({\bf S})\rangle_{\lambda,x g}\Big| \\
&&\leq N^{2\alpha-2} \beta \Big| \int_0^\mu dx |x|\frac{\partial}{\partial \lambda}{\mathbb E }  \langle h_N({\bf S})\rangle_{\lambda,x g}\Big|
\\&&\leq   N^{\alpha -1} \beta C_h \Big| \int_0^\mu dx\Big| 
= N^{\alpha-1}\beta  C_h |\mu|.  
\end{eqnarray}
We have used Lemma \ref{1}. The right hand side vanishes in the  infinite volume limit can be taken for $\alpha < 1$, and this completes the proof.
$\Box$

Lemma \ref{cont0} and Assumption 1 guarantee the existence of $p(\beta,\lambda, \mu)$ for each $(\beta,\lambda,\mu) \in (0,\infty) \times {\mathbb R}$.\\
Next, we prove an identity between expectation values of an arbitrary bounded operator for $\mu \neq 0$ and $\mu=0$ 
in a method  similar to that  used in Ref \cite{I3}.

{\lemma  \label{cont1} 
Let $f$ be a bounded function of spin operators whose infinite volume limit 
 \begin{equation}
\lim_{N\nearrow\infty}{\mathbb E}  \langle f ({\bf S}) \rangle_{\lambda, 0} 
 \end{equation} 
exists at $\mu=0$. Then the infinite volume limit 
% \begin{equation}\lim_{N\nearrow\infty}{\mathbb E}  \langle f ({\bf S}) \rangle_{\lambda, \mu g}  \begin{equation} 
at $\mu  \neq 0$ exists for  $\alpha < 1$  and for almost all  $\lambda \in {\mathbb R}$, and 
\begin{equation}
\lim_{N\nearrow\infty}{\mathbb E}\langle f ({\bf S}) \rangle_{\lambda,\mu g} 
=  \lim_{N\nearrow\infty}{\mathbb E}\langle f ({\bf S}) \rangle_{\lambda,0}.
\label{c1}
\end{equation}
% this function is identical to that  at $\mu=0$ 
 
 \noindent
Proof.} Integration of the derivative function  over the interval $(0,\mu)$ for an arbitrary $\mu \in{\mathbb R}$  gives
\begin{eqnarray}
&&{\mathbb E}\langle f ({\bf S}) \rangle_{\lambda,\mu g} 
-{\mathbb E}\langle f ({\bf S}) \rangle_{\lambda,0}
=\int_0^\mu dx \frac{\partial }{\partial x}{\mathbb E}\langle f ({\bf S}) \rangle_{\lambda,x g} 
\\
&&=\int_0^\mu dx {\mathbb E } N^{\alpha}\beta g (f({\bf S}); h_N({\bf S}) )_{\lambda,x g} 
\\
&&=\int_0^\mu dx {\mathbb E }N^{2\alpha}\beta^2 x (f({\bf S}); h_N({\bf S}) ; h_N({\bf S}))_{\lambda,x g} 
\\
&&=\int_0^\mu dxx N^{2(\alpha-1)}  \frac{\partial^2}{\partial \lambda^2}{\mathbb E }  \langle f({\bf S})\rangle_{\lambda,x g} .
\end{eqnarray}
Integration over an arbitrary interval of  $\lambda$ and Lemma \ref{1} imply
\begin{eqnarray}
&&\Big|\int_a^b d \lambda [{\mathbb E}\langle f ({\bf S}) \rangle_{\lambda,\mu g} -{\mathbb E}\langle f ({\bf S}) \rangle_{\lambda,0}]\Big|
\\
&&=\Big| \int_0^\mu dxx N^{2(\alpha-1)} \Big[ \frac{\partial}{\partial b}{\mathbb E }
\langle f({\bf S})\rangle_{b,x g} - \frac{\partial}{\partial a}{\mathbb E } 
\langle f({\bf S})\rangle_{a,x g} \Big]  \Big|\\
&&\leq \Big| \int_0^\mu dx|x| N^{2(\alpha-1)} \Big[ \Big|\frac{\partial}{\partial b}{\mathbb E }
\langle f({\bf S})\rangle_{b,x g} \Big| +\Big| \frac{\partial}{\partial a}{\mathbb E } 
\langle f({\bf S})\rangle_{a,x g} \Big| \Big] \Big|\\
&&\leq2N^{\alpha-1}\Big| \int_0 ^\mu dx C_f \Big| = 2N^{\alpha-1} C_f |\mu|.
\end{eqnarray}
The right hand side vanishes in the infinite volume limit  for $\alpha <1$.
Since the integration interval $(a,b)$ is arbitrary, the integrand in  the left hand side vanishes
for almost all $\lambda$ in the infinite volume limit.  
This completes the proof. $\Box$\\

{\lemma \label{cont2} 
 The infinit volume limit of the following function
 \begin{equation}
 \lim_{N\nearrow\infty}{\mathbb E}\langle h_N ({\bf S}) \rangle_{\lambda,\mu g} ^2
 \end{equation}
  exists 
 for $\alpha < 1$ for almost all  $\lambda \in {\mathbb R}$, and it is identical to that at $\mu=0$ 
\begin{equation}
\lim_{N\nearrow\infty}{\mathbb E}\langle h_N ({\bf S}) \rangle_{\lambda,\mu g} ^2
=  \lim_{N\nearrow\infty}{\mathbb E}\langle h_N ({\bf S}) \rangle_{\lambda,0}^2.
\label{c2}
\end{equation}
\noindent Proof.}  Integration of the derivative function  over the interval $(0,\mu)$ for an arbitrary $\mu\in{\mathbb R}$ gives
\begin{eqnarray}
&&{\mathbb E}\langle h_N ({\bf S}) \rangle_{\lambda,\mu g} ^2
-{\mathbb E}\langle h_N ({\bf S}) \rangle_{\lambda,0}^2
=\int_0^\mu dx \frac{\partial }{\partial x}{\mathbb E}\langle h_N ({\bf S}) \rangle_{\lambda,x g} ^2\\
&&=2 \int_0^\mu dx {\mathbb E } N^{\alpha}\beta g (h_N({\bf S}); h_N({\bf S}) )_{\lambda,x g} \langle h_N ({\bf S}) \rangle_{\lambda,x g}
\\
&&=2 \int_0^\mu dx {\mathbb E }N^{\alpha}\beta \frac{\partial}{\partial g} (h_N({\bf S}); h_N({\bf S}) )_{\lambda,x g} \langle h_N ({\bf S}) \rangle_{\lambda,x g} \\
&&=2\beta  \int_0^\mu dxx N^{2\alpha-1}  \frac{\partial}{\partial \lambda}{\mathbb E } 
(h_N({\bf S}); h_N({\bf S}) )_{\lambda,x g} \langle h_N ({\bf S}) \rangle_{\lambda,x g} 
\end{eqnarray}
Integration over an arbitrary interval of  $\lambda$ and Lemma \ref{1} imply
\begin{eqnarray}
&&\Big|\int_a^b d \lambda [{\mathbb E}\langle h_N ({\bf S}) \rangle_{\lambda,\mu g} ^2-{\mathbb E}\langle h_N ({\bf S}) \rangle_{\lambda,0}^2] \Big|\\
&&=\Big|2 \beta \int_0^\mu dxx N^{2\alpha-1} \Big[{\mathbb E } 
(h_N({\bf S}); h_N({\bf S}) )_{b,x g}\langle h_N({\bf S})\rangle_{b,x g} 
\nonumber \\
&&- {\mathbb E } (h_N({\bf S}); h_N({\bf S}) )_{a,x g}
\langle h_N({\bf S})\rangle_{a,x g} \Big]\Big|\\
&&\leq2 \beta\Big| \int_0^\mu dx|x| N^{2\alpha-1} \Big[{\mathbb E } 
|(h_N({\bf S}); h_N({\bf S}) )_{b,x g}||\langle h_N({\bf S})\rangle_{b,x g} |
\nonumber \\
&&+ {\mathbb E }| (h_N({\bf S}); h_N({\bf S}) )_{a,x g}|
|\langle h_N({\bf S})\rangle_{a,x g} |\Big] \Big|\\
&&\leq2 \beta C_h \Big|\int_0^\mu dx|x| N^{2\alpha-1} \Big[{\mathbb E } 
(h_N({\bf S}); h_N({\bf S}) )_{b,x g}
+ {\mathbb E } (h_N({\bf S}); h_N({\bf S}) )_{a,x g}\Big]\Big| \\
&&=2 C_h \Big|\int_0^\mu dx|x| N^{2(\alpha-1)} \Big[ \frac{\partial}{\partial b}{\mathbb E } 
\langle h_N({\bf S})\rangle_{b,x g} +\frac{\partial}{\partial a}{\mathbb E } 
\langle h_N({\bf S})\rangle_{a,x g} \Big]\Big|\\
&&\leq 4 N^{\alpha-1}  \Big|\int_0 ^\mu dx C_h^2\Big| = 4N^{\alpha-1} C_h^2 |\mu|.
\end{eqnarray}
The right hand side vanishes in the infinite volume limit for $\alpha < 1$.
Since the integration interval $(a,b)$ is arbitrary, the integrand in the left hand side vanishes
for almost all $\lambda$   in the infinite volume limit.  This completes the proof. $\Box$\\

Now, we prove Theorem \ref{MT}.  Lemma \ref{1} indicates that
the artificial  random perturbation suppresses the variance of the corresponding perturbation operator
even in the weak coupling and it vanishes in the infinite volume limit.   
Lemmas \ref{cont1} and \ref{cont2} imply that  the variance at $\mu=0$ is identical to that at $\mu \neq 0$ 
for  almost all non-random perturbation $\lambda$  in the infinite volume limit. 
Assumption 3 is necessary to prove Theorem \ref{MT} for quantum systems.

\paragraph{Proof of }Theorem \ref{MT}\\
Lemma \ref{1} for $f({\bf S})=h_N({\bf S})$ and $k=1$ yields
\begin{equation}
{\mathbb E} (h_N({\bf S});h_N({\bf S}))_{\lambda,\mu g} \leq \frac{C_h}{\beta |\mu|} N^{-\alpha},
\label{lemma1}
\end{equation}
for an arbitrary $\lambda \in {\mathbb R}$ and $\mu \neq 0$.
Harris'  inequality of the Bogolyubov type
between the Duhamel product and the Gibbs expectation of the square of arbitrary self-adjoint operator $O$ \cite{H}
\begin{equation}
( O,  O)_{\lambda,\mu g} \leq \langle {O}^2\rangle_{\lambda,\mu g} \leq ( O,  O)_{\lambda,\mu g}  
+\frac{\beta}{12} \langle[ O, [H,O]] \rangle_{\lambda,\mu g},
\label{harris}
\end{equation}
and Assumption 3 enable us to obtain  
 \begin{equation}
\lim_{N\nearrow\infty} {\mathbb E} \langle h_N({\bf S})^2 \rangle_{\lambda,\mu g} 
=\lim_{N\nearrow\infty}  {\mathbb E} ( h_N({\bf S}), h_N({\bf S}))_{\lambda,\mu g}. 
 \end{equation}
This and the bound (\ref{lemma1}) for $\mu \neq 0$ imply
 \begin{equation}
\lim_{N\nearrow\infty} {\mathbb E} [\langle h_N({\bf S})^2\rangle_{\lambda,\mu g}  - \langle h_N({\bf S})\rangle_{\lambda,\mu g}^2]
=\lim_{N\nearrow\infty}  {\mathbb E} ( h_N({\bf S}); h_N({\bf S}))_{\lambda,\mu g} = 0.
 \end{equation}
 This is true also for $\mu=0$ by Lemma \ref{cont1} and Lemma \ref{cont2} with by a uniform bound on $h_N^2$
 \begin{equation}
\lim_{N\nearrow\infty} {\mathbb E} [\langle h_N({\bf S})^2\rangle_{\lambda,0}  - \langle h_N({\bf S})\rangle_{\lambda,0}^2]
%=\lim_{N\nearrow\infty}  {\mathbb E} ( h_N({\bf S}); h_N({\bf S}))_{\lambda,\mu g}
 = 0.
 \end{equation}
This and Lemma \ref{Delta} give 
\begin{equation}
\lim_{N\nearrow\infty}
 [  {\mathbb E} \langle {h_N({\bf S})}^2 \rangle_{\lambda,0}-({\mathbb E} \langle {h_N({\bf S})}\rangle_{\lambda,0})^2]=0.
\label{1112}
\end{equation}
 This completes
the proof of Theorem \ref{MT}. $\Box$\\

\section{Applications to several models}

\subsection{Random energy model}
Random energy model is a well known  simple model where replica symmetry breaking appears. 
This model contains only $(S_i^z)_{i \in V_N}$ with spin $S=\frac{1}{2}$.
In the definition of the unperturbed  Hamiltonian (\ref{unpert}), $C_N := {\cal P}(V_N)$ and the function  $\varphi$ is defined by
\begin{equation}
\varphi({\bf S}_X) := \prod_{i \in X} \delta_{2S_i^z,1} \prod_{j\in X^c}  \delta_{2S_j^z,-1}
\end{equation}
The possible state in {\cal H} is represented in a spin configuration 
 $\sigma=(\sigma_i)_{i\in V_N} \in \Sigma_N := \{1,-1 \}^{V_N}$, which is a
  sequence of eigenvalues of the operators $(2  S_i^z)_{i \in V_N}$. 
  The function $\varphi$ defines  a natural  bijection ${\cal P}(V_N) \to \Sigma_N$, such that $2S_i ^z=1$ for $i \in X$ and   $2S_i^z =-1$ for $i \in X^c$.
  We identify a subset $X$ and the corresponding  spin configuration $\sigma$ with this bijection.
 Let us represent the Hamiltonian in terms of spin configurations.
  Let $(J_X)_{X \in C_N}$ be  i.i.d. standard Gaussian random variables in the Hamiltonian (\ref{unpert}), and
 identify them to ${\bf J}=(J_\sigma)_{\sigma \in \Sigma_N}$.
The Hamiltonian defines a partition function  
\begin{equation}
Z_N(\beta, {\bf J}) := \sum_{\sigma \in \Sigma_N} \exp (-\beta H_N(\sigma) ),
\end{equation}
where the unperturbed  Hamiltonian on $V_N$ can be  written in 
   \begin{equation}
  H_N(\sigma) := -\sqrt{N} J_\sigma.
   \end{equation}
%Define another  function by \begin{equation}
%P_N(\beta) := {\mathbb E} \log Z_N(\beta, J).\end{equation}
%then it is proved that $P_N(\beta)$ is positive and
 % \begin{equation}P_{M+N}(\beta) = P_M(\beta)+P_N(\beta) \begin{equation}
  %for arbitrary positive integers $M,N$ and for an arbitrary $\beta \in (0,\infty)$.  
Consider a $n$-replicated random energy model whose state is given by $n$ spin configurations $(\sigma^1, \cdots, \sigma^n) \in \Sigma_N^n$. 
The Hamiltonian of this model is given by
 \begin{equation}
H_{N,n}(\sigma^1, \cdots, \sigma^n):=\sum_{a=1}^n H_N(\sigma^a).
  \end{equation} 
Here we attach index $V$ to the Hamiltonian on $V_N$ for later convenience. 
This Hamiltonian is invariant under a permutation $s$
 \begin{equation}
H_{N,n}(\sigma^{s(1)},\cdots, \sigma^{s(n)}) = H_N(\sigma^1, \cdots, \sigma^n),
  \end{equation}
where  $s: \{1,2, \cdots, n \} \rightarrow \{1,2, \cdots, n \}$ is an arbitrary  bijection.  This symmetry is replica symmetry.
To study the spontaneous replica symmetry breaking, consider the following symmetry breaking perturbation
\begin{equation}%H_I (\sigma^1,\sigma^2):= -[N \lambda + N^\alpha \mu (\sqrt{u} g+ \sqrt{1-u} g') ] h_N(\sigma^1,\sigma^2),
h_N(\sigma^1,\sigma^2) := \prod_{i\in V_N} \delta_{\sigma_i^1, \sigma_i^2} .
\label{pREM}\end{equation}
This order parameter becomes finite if and only if two replicated spin configurations $\sigma^1$ and $\sigma^2$ are identical $(\sigma_i ^1)_{i\in V_N}
= (\sigma_i^2)_{i \in V_N}$, otherwise it vanishes.  
Note the upper  bound for this operator 
 \begin{equation}
\| h_N(\sigma^1,\sigma^2) \| \leq 1.
  \end{equation}
The partition  function is defined by
\begin{equation}
Z_{N,n}(\beta, \lambda, {\bf J}):= \sum_{\sigma^1, \cdots, \sigma^n } \exp[ -\beta H_{N,n}(\sigma^1, \cdots, \sigma^n) +\beta N \lambda h_N(\sigma^1,\sigma^2)].
\end{equation}
Dfine
\begin{equation}
\psi_{N,n}(\beta,\lambda,{\bf J}):= \frac{1}{N} \log Z_{N,n}(\beta, \lambda, {\bf J}),
\label{REMH}
\end{equation}
and
\begin{equation}
p_{N,n} (\beta,\lambda) :={\mathbb E} \psi_{N,n} (\beta, \lambda),
\end{equation}
whose infinite volume limit is
\begin{equation}
p_n(\beta, \lambda) :=\lim_{N\nearrow\infty} p_{N,n}(\beta, \lambda).
\end{equation} 
Guerra  obtains the following explicit form  \cite{G}
\begin{equation}
p_n(\beta,\lambda)=\max\{n p_1(\beta,0), \  p_1(2\beta,0) + \beta \lambda+(n-2)p_1(\beta,0)   \},
\label{Guerra}
\end{equation}
where
\begin{eqnarray}
p_1(\beta,0) & = & 
\left\{ 
\begin{array}{ll}
  \,  \beta \sqrt{2\log 2}  \ \ & (\beta> \sqrt{2 \log 2})  \\
 \,  \beta^2/2+ \log 2 
 \ \ & (\beta \leq \sqrt{2 \log 2}).
\end{array}
\right. 
\end{eqnarray} 
This shows  Assumption 1.\\

Assumption 2 is proved in the following lemma. 
{\lemma \label{vREM}
The variance of $\psi_{N,n}(\beta,\lambda, {\bf J})$ vanishes in the model defined by (\ref{REMH})   for any positive integers $N$ and 
for any $(\beta, \lambda) \in (0,\infty) \times {\mathbb R}$
\\

\noindent
Proof. } Let $(J_{\sigma^a})_{\sigma\in \Sigma_N, a=1,\cdots, n}$ $(J'_{\sigma^a})_{\sigma\in \Sigma_N, a=1,\cdots, n}$be i.i.d. standard Gaussian random variables.
Define ${\cal J}(u)= ({\cal J}_{\sigma^a}(u))_{\sigma^a \in \Sigma_N}$ of $u\in [0,1]$ by
\begin{equation}
{\cal J}_{\sigma^a}(u) := \sqrt{u}  J_{\sigma^a} + \sqrt{1-u} J_{\sigma^a}'
\end{equation}
for each spin configuration $\sigma \in \Sigma_N$ and $a=1,2$.
Define a function $\gamma(u)$ by
\begin{equation}
\gamma(u):={\mathbb E} [{\mathbb E }'\psi_{N,n}(\beta, \lambda, {\cal J}(u)) ]^2,
\end{equation}
where ${\mathbb E}$ ${\mathbb E}'$ denote the expectation over ${\bf J}$ and ${\bf J}'$ respectively.
Its derivative is evaluated as 
\begin{eqnarray}
&&\gamma\ '(u) =\frac{\beta}{N}{\mathbb E} {\mathbb E}'\psi_{N,n} (\beta, \lambda, {\cal J}(u)) 
{\mathbb E}'\sum_{a=1}^n \sum_{\sigma^1,\cdots, \sigma^n\in\Sigma_N} \Big(\frac{ J_{\sigma^a}}{\sqrt{u}}-\frac{ J_{\sigma^a}}{\sqrt{1-u}}\Big)\frac{e^{-\beta H_{N,n}( \sigma^1,\cdots, \sigma^n)}}{Z_{N,n}(\beta,\lambda,{\cal J}(u))} \nonumber
 \\
&& =\frac{\beta}{\sqrt{N}}{\mathbb E}\sum_{a}^n\sum_{\sigma^1, \cdots, \sigma^n \in\Sigma_N} \Big[ \frac{ 1}{\sqrt{u}} \frac{\partial }{\partial J_{\sigma^a}}{\mathbb E}'\psi_{N,n} (\beta, \lambda, {\cal J}(u)) 
{\mathbb E}'\frac{e^{-\beta H_{N,n}( \sigma^1,\cdots,\sigma^n)}}{Z_{N,n}(\beta,\lambda,{\cal J}(u))} \nonumber \\
&&- {\mathbb E}'\psi_{N,n} (\beta, \lambda, {\cal J}(u)) {\mathbb E}'\frac{ 1}{\sqrt{1-u}}\frac{\partial }{\partial J'_{\sigma^a}}\frac{e^{-\beta H_{N,n}(\sigma^1,\cdots, \sigma^n)}}{Z_{N,n}(\beta,\lambda,{\cal J}(u))} \Big] \\
&&=\frac{\beta^2}{N}{\mathbb E}\sum_{\sigma^1, \cdots, \sigma^n, \tau^1, \cdots, \tau^n \in\Sigma_N}\sum_{1\leq a,b \leq n}
{ \mathbb E}'  \delta_{\sigma^a,\tau^b}
\frac{e^{-\beta H_{N,n}(\sigma^1,\cdots, \sigma^n) }}{Z_{N,n}(\beta,\lambda,{\cal J}(u))} {\mathbb E}' 
\frac{e^{-\beta H_{N,n}(\tau^1,\cdots,\tau^n) }}{Z_{N,n}(\beta,\lambda,{\cal J}(u))} \\
&& \leq \frac{\beta^2n^2}{N},
\end{eqnarray}
for any $u \in [0,1]$. 
Then the variance of $\psi_{N,n}(\beta,\lambda)$ is
\begin{equation}
{\mathbb E} \psi_{N,n}(\beta, \lambda)^2 -p_{N,n}(\beta, \lambda)^2 = \gamma(1) -\gamma(0) = \int_0^1
du \gamma\ '(u) \leq \frac{\beta^2n^2}{N},
\end{equation}
for any positive integers $N$ and 
for any $(\beta, \lambda) \in (0,\infty) \times {\mathbb R}$. $\Box$\\

 Assumption 3 is satisfied trivially, since the Hamiltonian of this model commutes with $h_N$.
The following corollary  for the perturbed random energy model is obtained from Theorem \ref{MT}.

{\corollary \label{tREM}
In the $n$ replicated  random energy model perturbed by the Hamiltonian (\ref{pREM}) 
 in the infinite volume limit, for almost all $\lambda \in {\mathbb R}$,
 the expectation of the perturbing operator takes the value 
 \begin{equation}
\lim_{N\nearrow\infty} {\mathbb E} \langle h_N(\sigma^1,\sigma^2)  \rangle_{\lambda}   = 0 \ {\rm or} \ 1.
  \end{equation}
Proof.} Note the relation 
 \begin{equation}
h_N(\sigma^1,\sigma^2) ^2 =h_N(\sigma^1,\sigma^2).
  \end{equation}
Then, Theorem \ref{MT} implies
 \begin{equation}
\lim_{N\nearrow\infty} {\mathbb E} \langle h_N(\sigma^1,\sigma^2) \rangle_{\lambda} (1-{\mathbb E} \langle h_N(\sigma^1,\sigma^2) \rangle_{\lambda}  )=0. 
  \end{equation}
Therefore, $\displaystyle \lim_{n\to\infty} {\mathbb E} \langle h_N(\sigma^1,\sigma^2) \rangle_{\lambda}$  takes the value either  0 or1. $\Box$\\

Note that this corollary is also true for an arbitrary projection operator satisfying
$
h_N^2 = h_N
$   
in other models.

It is well known that the observation of
$
\displaystyle \lim_{\lambda\searrow 0} \lim_{N\nearrow\infty}  {\mathbb E} \langle h_N(\sigma^1,\sigma^2) \rangle_{\lambda}=1
$ 
implies the spontaneous replica symmetry breaking.
The replica symmetry breaking is also detect by the replica symmetric Gibbs state.
If  the replica symmetric calculation shows  
 \begin{equation}
0< \lim_{N\nearrow\infty}{\mathbb E}\langle h_N(\sigma^1,\sigma^2)\rangle_{0} <1,
  \end{equation} 
then this implies the finite variance 
 \begin{equation}
\lim_{N\nearrow\infty} [{\mathbb E} \langle h_N(\sigma^1,\sigma^2)^2 \rangle_{0} -({\mathbb E} \langle h_N(\sigma^1,\sigma^2) \rangle_{0})^2  ]>0
  \end{equation}
which gives an instability of the replica symmetric Gibbs state due to the large fluctuation.
At the same time, this implies  the non-commutativity of limiting procedure  
 \begin{equation}
\lim_{N\nearrow\infty} \lim_{\lambda\searrow 0}{\mathbb E}\langle h_N(\sigma^1,\sigma^2)\rangle_{\lambda} \neq \lim_{\lambda\searrow 0} \lim_{N\nearrow\infty}{\mathbb E}\langle h_N(\sigma^1,\sigma^2)\rangle_{\lambda}.
  \end{equation}
This is a typical phenomenon in spontaneous symmetry breaking.

Here, we point out an agreement between Corollary \ref{tREM} and Guerra's result \cite{G}.
Guerra has studied  the replica symmetry breaking in the random energy model  as a spontaneous symmetry breaking phenomenon. 
For $\beta \leq \beta_c = \sqrt{2\log2}$ and for a sufficiently small $\lambda$, $p_n(\beta,\lambda) = p_n(\beta,0)$, then Guerra's formula (\ref{Guerra})  gives
\begin{equation}
\lim_{\lambda \to 0} \lim_{N \nearrow \infty} {\mathbb E} \langle h_N (\sigma^1,\sigma^2)\rangle_\lambda =\frac{1}{\beta}\frac{\partial p_n}{\partial \lambda}=0.
\end{equation}
For $\beta > \beta_c $  becomes 
\begin{eqnarray}
p_n(\beta,\lambda) & = & 
\left\{ 
\begin{array}{ll}
  \, n \beta \sqrt{\log 2} +\beta \lambda \ & (\lambda>0)  \\
 \, n \beta \sqrt{\log 2} 
 & (\lambda \leq 0) ,
\end{array}
\right. 
\end{eqnarray} 
which implies 
\begin{equation}
\lim_{\lambda \searrow 0} \lim_{N \nearrow \infty} {\mathbb E} \langle h_N (\sigma^1,\sigma^2)\rangle_\lambda =
\lim_{\lambda \searrow 0}\frac{1}{\beta}\frac{\partial p_n}{\partial \lambda}(\beta,\lambda)=1,
\label{rsb}
\end{equation}
and
\begin{equation}
\lim_{\lambda \nearrow 0} \lim_{N \nearrow \infty} {\mathbb E} \langle h_N (\sigma^1,\sigma^2)\rangle_\lambda =\lim_{\lambda \nearrow 0}\frac{1}{\beta}\frac{\partial p_n}{\partial \lambda}(\beta,\lambda)=0.
\end{equation}
Corollary \ref{tREM} agrees with these results.
The non-differentiability of $p_n(\beta,\lambda)$ at $\lambda=0$ is  pointed out also by Mukaida \cite{M}.
On the other hand,
Guerra's  replica symmetric calculation of  the order parameter shows
 \begin{equation}
 \lim_{N\nearrow\infty} \lim_{\lambda \to 0}   {\mathbb E} \langle h_N (\sigma^1,\sigma^2)\rangle_\lambda= 
\lim_{N\nearrow\infty} {\mathbb E} \langle h_N(\sigma^1,\sigma^2) \rangle_{0}=1-\frac{\beta_c}{\beta} < 1.
  \end{equation}
  These show the non-commutativity of two limiting procedures.  
  Since the finite variance of $h_N$ shows the instability of the replica symmetric Gibbs state, it is not realistic and 
  the replica symmetry breaking Gibbs state  with the vanishing variance should be realized. 
  In this case, the identity  (\ref{rsb}) implies that  two replicated spin configurations $\sigma^1$and  $\sigma^2$ are identical.

\subsection{Quantum Heisenberg model without disorder}

Here we study spontaneous symmetry breaking of SU(2) invariance in the antiferromagnetic quantum Heisenberg model without disorder.
Let $V_N$ be a hyper cubic lattice $V_N:=[1,L]^d\cap {\mathbb Z}^d$ and
bipartite,  namely there exist two subsets $A$ and $B$ of $V_N$ such that $V_N=A \cup B$ and $A \cap B=\phi$. 
The model Hamiltonian is defined by
\begin{equation}
H_N({\bf S}):= \sum_{i \in A, j \in B} \sum_{p=x,y,z} J_{i,j}S_i^p S_j^p,
\end{equation}
where $J_{i,j} \geq 0$ is short-ranged and translationally invariant, i.e. there exists $c\geq 1$ such that  $J_{i,j}=0$ for any $|i-j|> c$, and
 $J_{i+v, j+v} =J_{i,j}$ for 
any $i,j,v\in V_N$.
Consider an antiferromagnetic order operator as a perturbation operator 
\begin{equation}
h_N({\bf S}) := \frac{1}{N}\Big( \sum_{i\in A} S^z_i -\sum_{j\in B} S_j^z \Big).
\end{equation}
This operator is bounded by
$
\| h_N({\bf S})\| \leq  S.
$
Define a perturbed  Hamiltonian by
 \begin{equation}
H:=H_N({\bf S}) -%(N^\alpha \mu g+
N \lambda h_N({\bf S}).
  \end{equation}
In this model, Assumption 1 is proved  in a standard method to show the sequence $p_N(\beta,\lambda)$  for 
positive integers $N$ becomes Cauchy for 
any $(\beta,\lambda) \in (0,\infty) \times {\mathbb R}$. 
Assumption 2 is trivial for the model without disorder and  
Assumption 3 is obvious for short-range interactions.   

Theorem \ref{MT} 
$
\displaystyle \lim _{N\nearrow\infty}[ \langle h_N({\bf S})^2 \rangle_{\lambda} - \langle h_N({\bf S}) \rangle_{\lambda}^2]=0
$
and SU(2) invariance $\langle h_N({\bf S}) \rangle_{0}=0$ yield the following corollary.

{\corollary  \label{AF} If the SU(2) invariant Gibbs state of the antiferromagnetic quantum Heisenberg model has a long-range order 
 \begin{equation}
\lim_{N\nearrow\infty} \langle h_N({\bf S})^2 \rangle_{0} \neq 0,
  \end{equation}
then we have
 \begin{equation}
\lim_{\lambda\searrow 0} \lim _{N\nearrow\infty}[ \langle h_N({\bf S})^2 \rangle_{\lambda} - \langle h_N({\bf S}) \rangle_{\lambda}^2]
\neq \lim _{N\nearrow\infty} \lim_{\lambda\searrow 0}[ \langle h_N({\bf S})^2 \rangle_{\lambda} - \langle h_N({\bf S}) \rangle_{\lambda}^2].
  \end{equation} 
 }
\\

The right hand side is non-zero, since the long-range order of the Gibbs state and its SU(2) invariance recovered by taking the  limit $\lambda \to 0$ first.
On the other hand, Theorem \ref{MT} states that the left hand side vanishes. 
The non-commutativity of limiting procedures in Corollary \ref{AF} claims the spontaneous SU(2) symmetry breaking, when a long-range order exists. 
%Dyson, Lieb and  Simon conjectured
Koma and Tasaki have shown  that  the long-range order ( equivalent  to a
finite variance of order operator ) in the symmetric Gibbs state 
implies the spontaneous  symmetry breaking  in the quantum 
Heisenberg model  with short-range antiferromagnetic  interactions \cite{KT}. 
They have proved 
 \begin{equation}
\sqrt{\lim _{N\nearrow\infty}  \langle h_N({\bf S})^2 \rangle_{0}} \leq \lim_{\lambda\searrow 0} \lim _{N\nearrow\infty} \langle h_N({\bf S}) \rangle_{\lambda}.
  \end{equation}
For ferromagnetic case, the corresponding inequality was proved by Griffiths \cite{Gff}.
 Even though the symmetric Gibbs state with the long-range order is unstable and unrealistic, 
it is mathematically well defined and can detect the symmetry breaking in the evaluation result of the finite variance of order operator.
Recently, Tasaki has shown that  the variance of order operator vanishes in the symmetry breaking ground state in the infinite volume limit \cite{hT}.     
Theorem \ref{MT} for quantum spin systems is consistent with his result.

\subsection{Quantum Edwards-Anderson model}

It is quite interesting whether or not, a replica symmetry breaking occurs in short-range disordered spin systems
as in the Sherrington-Kirkpatrick  model described by the Parisi formula \cite{Pr,T2,T}. 
Here, we discuss a  replica symmetry breaking as a spontaneous symmetry breaking phenomenon
in the quantum Edwards-Anderson model \cite{EA}.

Let $(S_j^p)_{j\in V_N,p =x,y,z}$ be spin operators
on a $d$-dimensional hyper cubic lattice $V_N:=[1,L]^d \cap {\mathbb Z}^d$, where $N = |V_N|= L^d$.
Let  $A$ be  a bounded subset  of $V_N$, such that  $|A| \leq C$, where $C$ is a positive constant independent of $N$.   
Define a collection of interaction ranges by
\begin{equation}
C_N:= \{X  | X = A+ v  \subset V_N, v \in V_N  \}.
\label{IR}
\end{equation}
Define  
 \begin{equation}S_X^p := \prod_{j \in X} S_j^p
  \end{equation}
for $X \in C_N$ and for $p=x,y,z.$
The Hamiltonian has disordered short-range interaction 
\begin{equation}
H_N({\bf S}, {\bf J}):= -\sum_{X \in C_N} \sum_{p=x,y,z} J_XK^pS_X^p,
\end{equation}
 where $(J_X)_{X \in C_N%,p=x,y,z
 }$ are i.i.d standard  Gaussian random variables and positive constants $(K^p)_{p=x,y,z}$.
 The interaction is short-ranged and translationally invariant, where
 $|C_N| \leq |A| N$. 
 Consider a $n$-replicated model with spin operators 
 $
 (S_j^{p,1}, \cdots, S_j^{p,n} )_{j \in V_N, p=x,y,z}
 $ 
and define a replica symmetric Hamiltonian
\begin{equation}
\sum_{a=1} ^n H_N({\bf S}^a,{\bf J}).
\label{dQSM}
\end{equation}
Define  a spin overlap as a perturbing operator
\begin{equation}
h_N({\bf S}^1,{\bf S}^2) := \frac{1}{N} \sum_{i\in V_N} S^{z,1}_iS^{z,2}_i,
\label{RSBO}
\end{equation}
which  breaks the replica symmetry.  Note the following bound 
 \begin{equation}
\| h_N({\bf S}^1,{\bf S}^2) \| \leq S^2.
  \end{equation}
Consider  the model defined by 
\begin{equation}
H_{N,n} ({\bf S}^1, \cdots, {\bf S}^n,{\bf J}) := \sum_{a=1} ^n H_N({\bf S}^a,{\bf J})-
N \lambda h_N({\bf S}^1,{\bf S}^2).
\label{pdQSM}
\end{equation}
To consider replica symmetry breaking, we attach the index $n$ to several functions in this subsection as in Subsection 3.1.
In this model, Assumption 1 is proved  in a standard method to show the sequence $p_{N,n}(\beta,\lambda,0)$ for  positive integers $N$ becomes Cauchy for 
any $(\beta,\lambda,0) \in (0,\infty) \times {\mathbb R}$
as  in the previous model. 
Assumption 3 is obvious for short-range interactions.
Assumption 2 
 is proved in the following lemma. 
%{\lemma \label{IVL} The infinite volume limit  of $p_N(\beta, \lambda,0)$ defined by the Hamiltonian (\ref{pdQSM}) 
%exists for each $(\beta, \lambda ) \in (0,\infty) \times {\mathbb R}$.\\Proof. }
{\lemma \label{varpsi} 
The variance  of $\psi_{N,n}(\beta, \lambda,
{\bf J} )$ defined by the Hamiltonian (\ref{pdQSM}) 
vanishes in the infinite volume limit
 for each $(\beta, \lambda) \in (0,\infty) \times {\mathbb R}$.
 \\
 
 \noindent
Proof. }
%Define a generating function
To prove this, we employ the generating  function $\gamma(u)$.
Let ${\bf J}' := (J_X')_{X \in C_N}$ be i.i.d. standard Gaussian random variables, and 
define 
\begin{equation}
{\cal J}(u):= \sqrt{u}{\bf J}+ \sqrt{1-u} {\bf J}'
\end{equation}
with $u\in [0.1].$
Define a generating function
\begin{equation}
\gamma(u) := {\mathbb E} [{\mathbb E}' \psi_{N,n}(\beta,\lambda,{\cal J}(u))]^2,
\end{equation}
where  ${\mathbb E}'$ stands for the expectation over only ${\bf J}'$.
Its derivative in $u$ is evaluated in the integration by parts
\begin{eqnarray}
\gamma\ '(u) &=&{\mathbb E} {\mathbb E}' \psi_{N,n} (\beta,\lambda,{\cal J}(u))\sum_{X\in C_N}{\mathbb E}'\Big( \frac{J_X}{\sqrt{u}}- \frac{J_X'}{\sqrt{1-u}}\Big) \frac{\partial 
\psi_{N,n}}{\partial {\cal J}_X}
 \\
 &=& \sum_{X\in C_N} {\mathbb E}\Big[ \frac{1}{\sqrt{u}} \frac{\partial}{\partial J_X}{\mathbb E}' \psi_{N,n} (\beta,\lambda,{\cal J}(u))\frac{\partial \psi_{N,n}}{\partial {\cal J}_X}
 \\&&-  {\mathbb E}' \psi_{N,n} (\beta,\lambda,{\cal J}(u)){\mathbb E}'\frac{1}{\sqrt{1-u}} \frac{\partial }{\partial J_X'}\frac{\partial \psi_{N,n}}{\partial {\cal J}_X} \Big]
 \\
 &=&  \sum_{X\in C_N} {\mathbb E}\Big[ {\mathbb E}' \frac{\partial \psi_{N,n}}{\partial {\cal J}_X} \Big]^2
 \\
&=& \frac{\beta^2}{N^2}\sum_{X \in C_N} { \mathbb E} \Big( {\mathbb E}' \sum_{p=x,y,z} 
\sum_{a=1} ^n K^p \langle S_X ^{p,a} \rangle_u) \Big)^2 
 \\
&\leq&
 \frac{ \beta^2|A|n^2 S^{2|A|}}{ N}(\sum_{p=x,y,z} K^p)^2,
\end{eqnarray}
where  we denote 
$$
\langle f({\bf S}^1, \cdots, {\bf S}^n) \rangle_u := \frac{1}{Z_{N,n}(\beta,\lambda,{\cal J}(u))} {\rm Tr[} f({\bf S}^1, \cdots, {\bf S}^n) 
e^{-\beta H_{N,n}({\bf S}^1, \cdots, {\bf S}^n, {\cal J}(u))}],
$$
for the Gibbs expectation of an arbitrary function $f$ of spin operators.
The variance of $\psi_N$ is given by
 \begin{equation}
 {\mathbb E} \psi_{N,n}(\beta,\lambda)^2 -p_{N,n}(\beta,\lambda)^2= \int_0^1 du \gamma\ '(u) \leq   \frac{\beta^2 |A| n^2S^{2|A|} }{ N}(\sum_{p=x,y,z} K^p)^2.
 \end{equation}
Then the variance of $\psi_{N,n}(\beta,\lambda)$ vanishes in the infinite volume limit  for arbitrary $(\beta,\lambda) \in (0,\infty) \times {\mathbb R}$.
$\Box$ \\

Chatterjee's definition of  replica symmetry breaking  \cite{C2} is  the following  finite variance 
of spin overlap in the replica symmetric Gibbs state with $\lambda=\mu=0$
 \begin{equation}
\lim_{N\nearrow\infty} {\mathbb E} \langle(  h_N({\bf S}^1,{\bf S}^2)  - {\mathbb E}\langle  h_N ({\bf S}^1,{\bf S}^2)\rangle_{0})^2 \rangle_{0} > 0.
  \end{equation}
Theorem \ref{MT} gives
 \begin{equation}\lim_{\lambda\searrow 0} \lim _{N\nearrow\infty}{\mathbb E} \langle(  h_N({\bf S}^1,{\bf S}^2)  - {\mathbb E}\langle  h_N ({\bf S}^1,{\bf S}^2)\rangle_{\lambda})^2 \rangle_{\lambda}=0,
   \end{equation}
then this yields the following corollary.
 
 {\corollary  If the replica symmetry breaking defined by  Chatterjee occurs in the model defined by the Hamiltonian (\ref{dQSM}),
the following limiting procedures do not commute
 \begin{equation}
\lim_{N\nearrow\infty} \lim_{\lambda\searrow 0}{\mathbb E} \langle(  h_N({\bf S}^1,{\bf S}^2)  - {\mathbb E}\langle  h_N ({\bf S}^1,{\bf S}^2)\rangle_{\lambda})^2 \rangle_{\lambda}
\neq \lim_{\lambda\searrow 0} \lim _{N\nearrow\infty}{\mathbb E} \langle(  h_N({\bf S}^1,{\bf S}^2)  - {\mathbb E}\langle  h_N ({\bf S}^1,{\bf S}^2)\rangle_{\lambda})^2 \rangle_{\lambda}.
  \end{equation}   }
  
 Ref. \cite{I3} indicates that the variance of the order operator (\ref{RSBO}) vanishes  by the disordered replica symmetry breaking perturbation
  \begin{equation}
 \sum_{i\in V_N} (\nu g_i +\lambda)S_i^{z,1} S_i^{z,2},
   \end{equation} 
 with Gaussian random variables $g_i$ and  constants $(\lambda, \nu) \in {\mathbb R}$.  Even  for $\nu=0$,  however,  Theorem \ref{MT}  
 implies that the variance of the order operator (\ref{RSBO}) vanishes  for almost all $\lambda \in {\mathbb R} $.\\

 {\bf Acknowledgment}\\ It is pleasure  to thank H. Mukaida for helpful discussions on the random energy model.


\begin{thebibliography}{13}
\bibitem{AC} Aizenman, M.,  Contucci, P. :{On the stability of quenched state in mean-field spin glass models}. J. Stat. Phys. \textbf{92}, 765-783(1997)

%\bibitem{AGL} Aizenman, M., Greenblatt,R.L.,  Lebowitz, J. L. :{Proof of rounding by quenched disorder of first order transitions in low-dimensional quantum systems} J. Math. Phys. {\textbf 53} 10.1063, (2012)

%\bibitem{AW} Aizenman, M.  Wehr, J. :{Rounding effects of quenched randomness on first-order phase transitions}.  Commun. Math. Phys. \textbf{ 130}, 489-528(1990) 

%\bibitem{BCF} Borgs, C., Chayes, J. T., Fr\"olich, J. :{Dobrushin states in quantum lattice systems}. Commun. Math. Phys. \textbf{ 189} 591-619, (1997) 

\bibitem{C2} Chatterjee, S. :{Absence of replica symmetry breaking in the random field Ising model}. Commun. Math .Phys. \textbf{337}, 93-102(2015)

%\bibitem{C1} Chatterjee,S.:{ The Ghirlanda-Guerra identities without averaging}. preprint, arXiv:0911.4520 (2009).

\bibitem{C} Chatterjee, S. :{Disorder chaos and multiple valleys in spin glasses}. preprint, arXiv:0907.3381 (2009).  

%\bibitem{CG} Contucci, P., Giardin\`a, C. :{ Spin-glass stochastic stability: A rigorous proof}. Annales Henri Poincare, \textbf{ 6}, 915-923, (2005)  

\bibitem{CG2} Contucci, P., Giardin\`a, C. :{ The Ghirlanda-Guerra identities}. J. Stat. Phys. \textbf{ 126}, 917-931,(2007)

%\bibitem{CG3} Contucci, P., Giardin\`a, C. :{ Perspectives on spin glasses.} Cambridge university press, 2012.

%\bibitem{CGP} Contucci, P.,  Giardin\`a, C., Pul\'e, J. :{ The infinite volume limit for finite dimensional classical and quantum disordered systems}. Rev.  Math. Phys. \textbf{ 16}, 629-638, (2004)

%\bibitem{CK} Campanino, M., Klein, A. :{Decay of Two-Point Functions for (d + 1)-Dimensional Percolation, Ising and Potts Models with d-Dimensional Disorder}. Commun. Math .Phys. \textbf{135}, 483-497(1991)

%\bibitem{CKP} Campanino, M., Klein, A., Pelez, J. F.,  :{Localization in the Ground State of the Ising Model with a Random Transverse Field}. Commun. Math. Phys. 135, 499-515 (1991)

%\bibitem{CL} Contucci, P., Lebowitz, J. L. :{ Correlation inequalities for quantum spin systems with quenched centered disorder}. J. Math. Phys. \textbf{ 51}, 023302-1 -6 (2010)

\bibitem{Cr} Crawford, N. :{ Thermodynamics and universality for mean field quantum spin glasses.} Commun. Math. Phys.\textbf{274}, 821-839(2007) 

%\bibitem{DLS} Dyson, Lieb, Simon,

 \bibitem{EA}  Edwards,S. F., Anderson, P. W. :{Theory of spin glasses} J. Phys. F: Metal Phys. \textbf{5}, 965-974(1975)


%\bibitem{FKG}  Fortuin,C. M., Kasteleyn P. W.,  Ginibre, J.:{ Correlation inequalities on some partially ordered sets}.Commun. Math. Phys. \textbf{22}, 89-103(1971). 



 \bibitem{GG} Ghirlanda, S., Guerra, F. :{General properties of overlap probability distributions in disordered spin systems. Towards Parisi ultrametricity}.
 J. Phys. A\textbf{31}, 9149-9155(1998)

\bibitem{GUW} Goldschmidt, C., Ueltschi, D., Windridge, P:{Quantum Heisenberg models and their probabilistic representations} Entropy and the quantum II, Contemp. Math. {\textbf 562} 177-224, (2011) 

%\bibitem{GAL} Greenbatt,R.L., Aizenman, M. Lebowitz, J. L. :{Rounding first order transitions in low-dimensional quantum systems with quenched disorder} Phys. Rev.Lett. {\textbf 103}19721(2009)

\bibitem{Gff} Griffiths, R. B. :{Spontaneous Magnetization in Idealized Ferromagnets} %Spontaneous magnetization in idealized ferromagnets}. Phys.Rev.  \textbf{152 }, 240-246, (1964)
Phys.Rev.{\textbf 152} 240-246 (1966)
 
\bibitem{G} Guerra,  F.:{The phenomenon of spontaneous replica symmetry breaking in complex statistical mechanics systems}J. Phys: Conf. Series \textbf{442} 012013(2013)

\bibitem{G3} Guerra,  F.:{Spontaneous replica symmetry breaking and interpolation methods for complex statistical mechanical systems} Lecture Notes in Mathematics \textbf{ 2143}, 45-70, Springer  (2013)

\bibitem{GT} Guerra,  F.   Tninelli, F. L.  : {Infinite Volume Limit and Spontaneous Replica Symmetry Breaking in Mean Field Spin Glass Models }  Ann. Henri Poincar'e  {\textbf 4}, Suppl. 1  S441-S444  (2003) 

\bibitem{GT2} Guerra,  F.  Tninelli, F. L. :{The infinite volume limit in generalized mean field disordered models} Markov Proc. Rel. Fields 9:2, 195-207 (2003).

%\bibitem{NS} Newman,  C. M., Stein, D. L.: {Non-mean-field behavior of realistic spin glasses}.Phys. Rev. Lett. \textbf{76}515-518 (1996)

\bibitem{H}  Harris, A.B. :{Bounds for certain thermodynamic averages}J. Math. Phys. 8 1044-1045.(1967) 

%\bibitem{AT}J. R. L. de Almeida and D. J.  Thouless ``Stability of the Sherrington-Kirkpatrick solution of spin glass model" J. Phys. A :Math.Gen. \textbf{ 11}(1978), 983-990

\bibitem{I} Itoi, C. :{General properties of overlap operators in disordered quantum spin systems} J. Stat. phys. {\textbf 163} 1339-1349 (2016) 
 %\bibitem{M} Mukaida, H. :{Generating function for the connected correlator in the random energy model and its effective potential} aXiv:1404.3130[cond-mat.dis-nn]

\bibitem{I2} Itoi, C. :{Universal nature  of replica symmetry breaking  in quantum systems with Gaussian disorder} J. Stat. phys. {\textbf 167} 1262-1279 (2017) 
 
\bibitem{I3}  Itoi, C. :{Absence of replica symmetry breaking in transverse and longitudinal  random field  Ising model} J. Stat. phys. {\textbf 170 } 684-699 (2018) 

\bibitem{KT} Koma, T., Tasaki, H. {Symmetry breaking in Heisenberg antiferromagnets} Commun. Math. Phys. {\bf 158}, 198-214 (1993).
 
 \bibitem{M2} Mukaida, H. :{Non-differentiability of the effective potential and the replica symmetry breaking in the random energy model} J. Phys. A \textbf{49} 45002,1-15 (2016)


%\bibitem{T2} F. L. Toninelli,  ``About the Almeida-Thouless transition line in the Sherrington-Kirkpatrick mean-field spin glass model" Europhysics Letters \textbf{ 60}(2002), 764-767
%\bibitem{NS1} H. Nishimori and D. Sherrington, AIP Conference Proceedings 553, 67 (2001)

%\bibitem{N} H. Nishimori, ``Statistical Physics of Spin Glasses and Information Processing: An Introduction"Oxford university press (2001)

\bibitem{Pn} Panchenko, D. :{ The Ghirlanda-Guerra identities for mixed $p$-spin glass model}. Compt. Read. Math. \textbf{ 348}, 189-192(2010).
 %The Parisi formula for mixed $p$-spin models" The Ann. of Probab. \textbf{ 42} No3 (2014), 946-958.  
 
%\bibitem{Pa} Panchenko, D. :{ A connection between the Ghirlanda-Guerra identities and ultrametricity}. Ann.  Prob.  \textbf{ 38}, 327-347, (2010) 

%\bibitem{Pa2} Panchenko, D. :{ The Parisi ultrametricity conjecture}. Ann.  Math. \textbf{ 177}, 383-393, (2013)

\bibitem{Pr} Parisi, G. :{A sequence of approximate solutions to the S-K model for spin glasses}. J. Phys. A {\textbf 13}, L-115 (1980)

\bibitem{S} Seiler, E., Simon, B. :{ Nelson's symmetry and all that in Yukawa and $(\phi^4)_3$ theories}. Ann.  Phys. \textbf{ 97}, 470-518, (1976)



%\bibitem{SK} Sherrington, S., Kirkpatrick, S :{ Solvable model of spin glass}. Phys. Rev. Lett. \textbf{ 35},  1792-1796, (1975). 
 
%\bibitem{Suzuki}  M. Suzuki, :{Relationship between d-dimensional quantal spin systems and (d+1)-dimensional Ising systems}. Prog. Theor. Phys. {\textbf 56}, 1454-1468 (1976)


\bibitem{T2}  Talagrand,  M. :{The Parisi formula}. Ann. Math. \textbf{ 163}, 221-263 (2006).

\bibitem{T} Talagrand, M. :{Mean field models for spin glasses}. Springer, Berlin (2011).
	 
\bibitem{hT} Tasaki, H. :{Long-range order, “tower” of states, and symmetry breaking in lattice quantum systems}  J. Stat. Phys. {\bf 174}, 735-761 (2019).

%\bibitem{C} W. Chen, ``On the mixed even-spin Sherrington-Kirkpatrick model with ferromagnetic interaction" Ann. Inst. Henri Poincare {\bf 50}(2014), 63-83.

%\bibitem{G} F. Guerra,  ``Broken replica symmetry bounds in the mean field spin glass model" Commun. Math. Phys. {\bf 233}(2003), 1-12.


%\bibitem{I} M. Itakura


 

%\bibitem{ASS} M. Aizenman, R. Sims and S. L. Starr``An extended variational principle for the SK spin-glass model" Phys. Rev. B {\bf 68} (2003) 214403.

 %\bibitem{N} H. Nishimori
 
%\bibitem{H} K. Hukushima
 
% \bibitem{MY} Mezard and Young

\end{thebibliography}
\end{document}